\documentclass[twocolumn]{emulateapj}
\usepackage{apjfonts}
\usepackage{epsfig}
\usepackage{hyperref}

\begin{document}
\def\gapprox{\mathrel{\vcenter{\offinterlineskip \hbox{$>$}
    \kern 0.3ex \hbox{$\sim$}}}}
\def\lapprox{\mathrel{\vcenter{\offinterlineskip \hbox{$<$}
    \kern 0.3ex \hbox{$\sim$}}}}

\newcommand{\Dt}[0]{\bigtriangleup t}
\newcommand{\Dx}[0]{\bigtriangleup x}
\newcommand{\E}{\mathcal{E}}
\newcommand{\D}{\bigtriangleup}
\newcommand{\beq}{\begin{equation}}
\newcommand{\eeq}{\end{equation}}
\newcommand{\mm}[2]{\textrm{minmod}\left({#1},{#2}\right)}
\newcommand{\sign}{\textrm{sign}}
\newcommand{\nf}{\mathcal{F}}

\shortauthors{Choi \& Stone}
\shorttitle{Effect of Anisotropic Conduction on TI}

\title{The Effect of Anisotropic Conduction on the Thermal Instability
       \\ in the Interstellar Medium}

\author{Ena Choi \& James M. Stone}
\affil{Department of Astrophysical Sciences, Princeton University, Princeton,
NJ 08544}

\begin{abstract}

Thermal instability (TI) can strongly affect the structure and dynamics of
the interstellar medium (ISM) in the Milky Way and other disk galaxies.
Thermal conduction plays an important role in the TI by stabilizing small 
scales and limiting the size of the smallest condensates.  In the 
magnetized ISM, however, heat is conducted anisotropically (primarily
along magnetic field lines).  We investigate the effects of anisotropic
thermal conduction on the nonlinear regime of the TI by performing
two-dimensional magnetohydrodynamic simulations.  We present 
models with magnetic fields of different initial geometries and strengths,
and compare them to hydrodynamic models with isotropic conduction.
We find anisotropic conduction does not significantly alter the overall
density and temperature statistics in the saturated state of the TI.
However, it can strongly affect the shapes and sizes of cold clouds
formed by the TI.  For example, for uniform initial fields long filaments
of cold gas are produced that are reminiscent of some observed HI 
clouds. For initially tangled fields, such filaments are not produced.  We 
also show that anisotropic conduction suppresses turbulence generated
by evaporative flows from the surfaces of cold blobs, which may have
implications for mechanisms for driving turbulence in the ISM.

\end{abstract}
\keywords{galaxies: ISM -- instabilities -- ISM: kinematics and dynamics}

\section{Introduction}\label{sec:Intro}

The thermal instability (TI) plays an important role in controlling several 
different aspects of the interstellar medium (ISM) and star formation. For 
example, it has been invoked to explain the observed multiphase 
structure of the ISM \citep{par53,spi58,fie65,fie69}. The linear stage of 
the TI in astrophysical gases was first studied in detail by \citet{fie65}. He 
identified three unstable modes: an isobaric and two isentropic modes.  
In the nonlinear regime, the isobaric mode produces condensations that 
are fundamental to the classical two-phase model of the ISM \citep{fie69}, 
as well as the extension to a three-phase model by \citet{cox74,mck77}. 
The TI also regulates the mass flow between the different components of 
the ISM, and therefore affects the star formation rate \citep{chi87}. Thus, it 
is important to investigate the role of TI in determining the distribution of 
density, temperature, and other physical variables in the multiphase ISM.
For this reason, a variety of authors have studied the linear and nonlinear 
stages of the TI using numerical hydrodynamic simulations
\citep{hen99,kri02a,kri02b,kri04,vaz00,gaz01,san02,pio04,pio05,koy04,vaz07,kim08,ino08}.

Thermal conduction is important to include in studies of the TI for two
reasons.  First, it suppresses the growth rate at small scales, in fact 
isobaric perturbations with wavelength smaller than the Field length 
$\lambda_F$ \citep{fie65} do not grow at all.  Second, it produces 
evaporation from the surfaces of cold dense fragments, and the 
interaction of the evaporative flows can induce turbulence.

Including thermal conduction is essential for numerical studies of the TI. 
Without explicit thermal conduction, perturbations at the grid scale grow 
fastest, and may eventually come to dominate the dynamics. For this 
reason, \citet{koy04} concluded that numerical studies of the TI must 
satisfy a "Field criterion", that is the Field length must be resolved by at 
least a few cells to prevent artificial fragmentation at the grid scale, and 
to avoid the results being dominated by grid noise.  Satisfying the Field 
criterion requires including explicit thermal conduction 
\citep{pio04,pio05,koy04,bra07}.  To highlight one example, 
\citet{pio04,pio05} studied the interaction of the TI and the 
magnetorotational instability (MRI) in disks including isotropic thermal
conduction; they found the MRI could drive turbulence and 
fragmentation in the diffuse ISM at amplitudes consistent with HI 
observations.

To date most studies of the TI with conduction have assumed the
conductivity is isotropic. However, in a magnetized plasma, electrons 
can flow more freely along magnetic field lines than across them, leading 
to anisotropic transport coefficients \citep{spi62}.  The degree of 
anisotropy is measured by the ratio of the electron gyro radius to the 
mean free path between collisions.  For the warm medium, 
where typically $T=1500$ K, $n=2$ cm$^3$, and $B=1 \mu$G, the 
Columb mean free path is $\lambda_{\rm mfp} \sim 10^{10}$ cm, while 
the electron gyro radius is $r_g=10^{6}$ cm.  Thus, in this medium the 
thermal conduction should be highly anisotropic, and primarily along 
magnetic field lines.

The implications of anisotropic transport terms on the dynamics of 
astrophysical plasmas has begun to be explored recently 
\citep{bal01,sha03,par05}.  For example, in stratified atmospheres,
anisotropic conduction can result in the magnetothermal and heat-flux 
buoyancy instabilities in the intracluster medium \citep{par08,sha09}.
In addition, it can have effects on the evolution of supernova remnants 
\citep{bal08} and on magnetized spherical accretion flows \citep{sha08}.
Recently, \citet{sha10} studied the TI in the intracluster medium, 
including heating by cosmic rays.  Their primary result is that with 
anisotropic thermal conduction, the TI could produce filaments of cold 
gas orientated along magnetic field lines.

In this study, we perform two-dimensional numerical hydrodynamic
and magnetohydrodynamic simulations of TI in the ISM that include the
 anisotropic heat conduction.  The purpose of this paper is to 
investigate the effects of anisotropic thermal conduction on the structure 
and dynamics of interstellar medium, and to compare the results to 
models which include only isotropic conduction. Since the geometry and 
strength of magnetic field in different regions of the ISM may vary, we 
perform simulations of the TI with various magnetic field strengths and 
two different initial geometries for the field. 

Ideally, the numerical studies of the thermal instability should include 
both isotropic and anisotropic conduction with the temperature 
dependency since the thermal instability develops density distribution 
with distinct peaks in the cold/dense phase dominated by isotropic 
conduction and in the diffuse/hot phase with higher ionization fraction 
and thus more affected by anisotropic conduction. However, including 
both isotropic and anisotropic conduction with a realistic temperature 
dependency is challenging, since it would decrease the Field length, and 
require higher resolution. In this study, we restrict our exploration to the 
simulations only with isotropic conduction or ones only with anisotropic 
conduction to better understand the effects of anisotropic conduction.

This paper is organized as follows: our numerical methods and code 
tests are summarized in \S~\ref{sec:Model}. In \S~\ref{sec:iso}, we first 
discuss the effects of the conduction rate and resolution on the TI in 
hydrodynamical simulations, and then use these results to choose a 
specific set of model parameters for our simulations. Results from 
calculations of the nonlinear regime of the TI with anisotropic conduction 
are presented in \S~\ref{sec:aniso}. We summarize and discuss our 
results in \S~\ref{sec:Summary}.

\section{Numerical Model}\label{sec:Model}
We solve the equations of ideal MHD with additional terms for radiative
cooling, heating, and heat conduction
\begin{eqnarray}
\frac{\partial \rho}{\partial t} +
{\bf\nabla\cdot} [\rho{\bf v}] & = & 0,
\label{eq:cons_mass} \\
\frac{\partial \rho {\bf v}}{\partial t} +
{\bf\nabla\cdot} \left[\rho{\bf vv} - {\bf BB} +
\left( p + \frac{B^2}{2}  \right) {\bf I} - {\bf \sigma} \right] & = & 0,
\label{eq:cons_momentum} \\
\frac{\partial E}{\partial t} +
\nabla\cdot \left[\left(E + p + \frac{B^2}{2}\right) {\bf v} - {\bf B} 
({\bf B \cdot v})+{\bf Q} - \bf v\cdot{\bf \sigma}\right] 
& = & - \rho\mathcal{L} ,
\label{eq:cons_energy} \\
\frac{\partial {\bf B}}{\partial t} -
{\bf\nabla} \times \left({\bf v} \times {\bf B}\right) & = & 0,
\label{eq:induction}
\end{eqnarray}
where the symbols have their usual meaning. The total energy density 
$E$ is given as
\begin{equation}
  E = \frac{P}{\gamma -1} + \frac{1}{2}\rho v^{2} + \frac{B^{2}}{2},
\label{eq:total_energy}
\end{equation}
where $\gamma$ is the ratio of specific heats (assumed to be $5/3$
throughout this paper). Here, ${\bf \sigma}$ is the viscous stress tensor 
\begin{equation}
\sigma_{ij} = \eta \left[ \left(\frac{\partial {\bf v}_i}{\partial x_j} + 
\frac{\partial {\bf v}_j}{\partial x_i} \right)
- \frac{2}{3} \delta_{ij} {\bf\nabla\cdot}{\bf v}  \right]
\end{equation}
where $\eta$ is the coefficient of viscosity and summation over 
repeated indices is implied. In equation \ref{eq:cons_energy}, ${\bf Q}$ 
denotes the heat flux and $\mathcal{L} = \rho\Lambda(\rho,T)-\Gamma$ 
is the net cooling function, where $\Lambda(\rho,T)$ and $\Gamma$ are 
the cooling and heating rates per unit mass respectively. These 
equations are written in units such that the magnetic permeability 
$\mu_B=1$.

The heating and cooling rates are 
\begin{equation}
\mu m_{\rm H}\Gamma = 2\times 10^{-26}\;{\rm erg\;s}^{-1},
\label{eq:E_heat}
\end{equation}
\begin{eqnarray}
\mu m_{\rm H}\frac{\Lambda(T)}{\Gamma}
     &= &10^7 \exp\left(\frac{-1.184\times 10^5}{T+1000}\right)\nonumber\\*
     &+ &1.4\times 10^{-2} \sqrt{T} \exp\left(\frac{-92}{T}\right)\;{\rm cm}^3\,
\label{eq:E_cool}
\end{eqnarray}
where we have adopted the functional fit to the ISM cooling rates as
suggested by \citet{koy02}\footnote{Equation \ref{eq:E_cool} contains 
the corrections pointed out by \citet{nag06}}.  For the net cooling curve 
adopted in this work, the transition temperatures that define the warm 
(phase "F", $T>T_{\rm max}$),  intermediate 
(phase "G", $T_{\rm min}<T<T_{\rm max}$), and cold (phase "H", at 
$T < T_{\rm min}$) phases are $T_{\rm max} = 5012$ K and 
$T_{\rm min} = 185$ K respectively.

In the case of {\em isotropic} conduction, the heat flux ${\bf Q}$ is given 
by
\begin{equation}
{\bf Q} = {\bf Q}_{\rm iso} = - \mathcal{K}{\nabla}T,
\label{eq:heat_aniso}
\end{equation}
while in the case of {\em anisotropic} conduction, the heat flux is given 
by
\begin{equation}
{\bf Q} = {\bf Q}_{\rm aniso} = - \mathcal{K}\mathbf{\hat{b}\hat{b}}\cdot\mathbf{\nabla}T.
\label{eq:heat_aniso}
\end{equation}
where $\mathbf{\hat{b}}$ is a unit vector in the
direction of the magnetic field, and $\mathcal{K}$ is the conductivity.
For the temperature ranges considered here, the latter is given by
\citep{par53, spi62}
\begin{equation}
\mathcal{K}=2.5 \times 10^3 \ T^{1/2} \rm{erg \ cm^{-1} \ K^{-1} \ s^{-1}}.
\label{eq:par}
\end{equation}

The relation between the dynamic viscosity coefficient $\eta$ and the 
thermal conductivity $\mathcal{K}$ can be characterized by the Prandtl 
number:
\begin{equation}
{\rm Pr}=\frac{\gamma}{\gamma-1}\frac{k_{\rm B}}{m_{\rm H}}\frac{\eta}{\mathcal{K}}.
\end{equation}
The Prandtl number for a neutral monoatomic gas is Pr=2/3.  For most of
the simulations presented in this paper, we include explicit viscosity with 
a constant coefficient of viscosity that corresponds to a Prandtl number 
$\rm Pr = 2/3$.

All of the numerical results presented in this paper were calculated using
Athena.  Details of the algorithms implemented in Athena are 
documented in \citet{gar05,gar08} and extensive tests of the MHD 
algorithms in Athena are shown in \citet{sto08}. In this work, we use the
Roe approximate Riemann solver and the directionally unsplit CTU 
(corner transport upwind) integration scheme.  Optically thin cooling was 
added to the integrators without using operator splitting.  To circumvent 
timestep constraints in regions of strong cooling, we limit the minimum 
temperature to the equilibrium value where heating balances cooling.  
This allows large time steps while avoiding the large errors or overshoots 
encountered with backward Euler or Crank-Nicholson implicit differencing.

We solve the equations in a rectangular domain with periodic boundary 
conditions and no gravity.  Our domain spans a region of size 
$2L \times L$, where $L$ is varied between different calculations.
We adopt an initial pressure of $P/k=3000$ ${\rm K~ cm^{-3}}$ 
and density of $n=2.0$ ${\rm cm^{-3}}$. The equilibrium pressure 
corresponding to this density is $P/k=2896.6$ ${\rm K~ cm^{-3}}$, so our 
initial condition is in a slowly cooling state. The mean mass per particle in 
units of the mass of atomic hydrogen is $\mu=1.27$, representing 10\% 
He abundance by number. To initiate the thermal instability (TI), we add 
random pressure perturbations with a maximum amplitude of 0.1\%. For 
our assumed initial temperature, the fiducial value of the conductivity is
\begin{equation}
\mathcal{K}_0=9.68 \times 10^4 \rm{erg \ cm^{-1} \ K^{-1} \ s^{-1}}
\end{equation}
which follows from equation \ref{eq:par}.

\begin{figure}
\epsfig{file=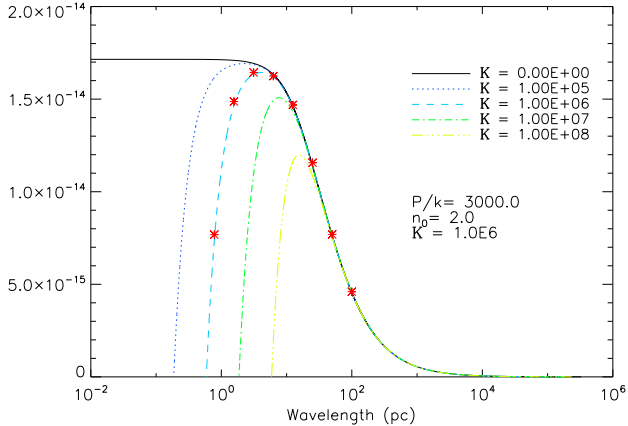,width=\columnwidth}
\caption{Numerically measured growth rates of the TI for different wave 
numbers and with a constant conductivity 
$\mathcal{K}= 10^6$ $\rm{ erg \ cm^{-1} \ K^{-1} \ s^{-1}}$ plotted as points,
overlaid with the theoretical growth rates for conductivities of
$\mathcal{K}= [0.0, 10^5,10^6,10^7,10^8] $ $\rm{ erg \ cm^{-1} \ K^{-1} \ s^{-1}}$.
The excellent agreement between the numerical and analytic growth 
rates verifies our numerical algorithms.
\label{fig1}}
\end{figure}

In order to test the implementation of the heating, cooling and 
conduction terms in our code, we have performed one-dimensional 
simulations of the TI, and compared the numerically measured growth 
rates with the theoretical prediction \citet{fie65}.  The initial conditions 
for this test are a medium at rest with constant density and pressure of
$n=2.0~{\rm cm^{-3}}$ and $P/k=3000$ ${\rm K~ cm^{-3}}$, and 
isotropic conduction with coefficient $\mathcal{K}= {\rm 10^6 erg~s^{-1} cm^{-1}
K^{-1}}$. The box size was $L=100$ pc and the grid contained 1024 
zones. We initialize the models with eigen modes of the instability by 
imposing sinusoidal fluctuations of amplitude 1 percent and with wave 
number $k$, and measure the growth rate of the density for different 
values of $k$. Figure \ref{fig1} compares the numerical growth rates from 
our simulations along with the theoretical values using four values of the 
thermal conductivity $\mathcal{K}$.  Clearly there is good agreement 
between the analytic and numerical growth rates; the numerical method 
reproduces the theoretical value of the growth rate to better than 4 percent 
in all cases.

For any non-zero value of the conductivity,  the thermal diffusivity becomes
comparable to the heating and cooling term at a critical wavelength 
referred to as the Field length, given approximately by \citet{fie65}
\begin{equation}
\lambda_F=2\pi\sqrt{\frac{\mathcal{K}T}{\rho^2\Lambda}}.
\label{eq:FieldLength}
\end{equation}
In the case of no conduction $\mathcal{K}=0$, the growth rate of the TI
is largest at the smallest scales ($\lambda=0$). For this reason, 
\citet{koy04} have shown that the inclusion of explicit thermal conduction 
is necessary in studies of the TI in order to damp growth at the grid scale 
and ensure the fastest growing modes of the TI \citet{fie65} are resolved 
by at least three cells.  Without explicit conduction, grid noise will be
amplified until it dominates the solution.  We further investigate the
consequences of simulating the TI without explicit thermal conduction
in the next section.

\section{TI with isotropic thermal conduction}\label{sec:iso}

In the following subsections, we investigate the role of varying the
amplitude of the conductivity, size of the computational domain, and grid
resolution using models with isotropic conduction.  This allows us to
decide on an optimal set of model parameters before considering the
effect of anisotropic conduction in the next section.

\subsection{Effect of varying the conductivity}\label{sec:coneffect}
\begin{table}
   \begin{center}
   \caption{Models with isotropic conduction.}
   \vskip+0.5truecm
   \begin{tabular}{c|c|c|c|c|c}\hline\hline
    {\bf \footnotesize Conductivity} & $\lambda_{F}$$^{a}$/$\lambda_{F,min}$ 
    & Box size (pc) &  $n_{\lambda_{F}}$$^{b}$  & $\Delta x$(pc) & $n_{c}$$^{c}$/$n_{c,min}$  \\
\hline
       0.0  & \-- & $4.0\times2.0$ & \--  & 0.00195 & \-- \cr
    
     $\mathcal{K}_0 $ & 0.12 / 0.006 & $4.0\times2.0$ & 33.3  & 0.00195 & 61.5 / 3 \cr
     $4\mathcal{K}_0 $ & 0.24 /  0.012 & $8.0\times4.0$& 33.3  & 0.0039 & 61.5 / 3\cr
     $16\mathcal{K}_0$ & 0.48 / 0.024 & $16.0\times8.0$ & 33.3  & 0.0078 & 61.5 / 3\cr
     $64\mathcal{K}_0 $ & 0.96 / 0.048 & $32.0\times16.0$ & 33.3  & 0.0156 & 61.5 / 3 \cr
  \hline\hline
   \end{tabular}
   \end{center}
   \label{tab1}
$^{a}$Field length in parsec as given by Equation~\ref{eq:FieldLength}.\\
$^{b}$The number of Field lengths along the longest axis of the simulation box ($2L$).\\
$^{c}$Field number, i.e., the number of zones in a Field length.
 \end{table}

With our adopted initial condition of density and temperature and with
the standard level of thermal conduction $\mathcal{K}_0$, the Field
length $\lambda_{F}$ is about $~0.12$ pc.  Resolving this length in a
computational domain that spans hundreds of parsecs is very 
demanding. Thus, in most previous work, the thermal conductivity has 
either been increased (by about a factor of 100) in order to resolve the 
Field length \citep{pio04,pio05,bra07}, or the computational domain has 
been chosen to be very small (for example $0.3 \-- 4.8 \rm pc$) 
\citet{koy04,koy06} so that the standard value $\mathcal{K}_0$ can be 
used.  In this section, we study whether the properties of the nonlinear 
regime of the TI depend on the value of the conductivity adopted.

We present the results of five inviscid hydrodynamical models which
include isotropic conduction with values for the conductivity of
$\mathcal{K}=[0.0,\mathcal{K}_0,4\mathcal{K}_0,16\mathcal{K}_0,
64\mathcal{K}_0]$.   Table 1 summarizes the properties of the
simulations.

Each model begins from a uniform homogeneous medium at rest with 
the fiducial density and pressure.  Each uses the same numerical 
resolution, $2048 \times 1024$ cells, but the size of the domain $L$ is 
varied so that the Field length is resolved, even in the nonlinear regime.  
Since the Field length is proportional to $\sqrt{T}$ and inversely 
proportional to density $n$, as density inhomogeneities are amplified 
by the TI the local value of the Field length varies, i.e., the local Field 
length in the regions of hot gas increases and that in regions of cold 
gas decreases. By the end of the simulation, we find the Field length in 
the cold gas is $\sim 20$ times smaller than in the initial conditions. To 
fully resolve the Field length throughout the simulation, grid cells must 
be smaller than the Field length {\it in the cold phase.} We calculate the 
minimum value of the Field length $\lambda_{F,min}$ for cold gas and 
keep the Field number $n_{c}=\lambda_{F}/ \delta x$, greater than 3
throughout our simulations.

We choose the physical size of the domain $L$ to be large enough so 
that it spans a large number of Field lengths, similar to the value used in
previous work \citep{pio04, pio05, bra07}.  Defining $n_{\lambda_{F}}$
as the number of Field lengths along the longest axis of the simulation
box ($2L$), we vary $L$ so that for each value of the conductivity,
$n_{\lambda_{F}}$ is fixed. We have also performed a large number of 
simulations with the same conductivity $\mathcal{K}$ but different values 
of $L$ to confirm that the statistical properties of the density and the level 
of kinetic energy in the nonlinear stage of the  TI are not affected by 
changing $L$ (provided that $n_{\lambda_{F}}$ is large).  Thus, we are 
confident we can compare runs with different physical sizes to isolate the 
effect of varying the conductivity on the TI.

\begin{figure}
\epsfig{file=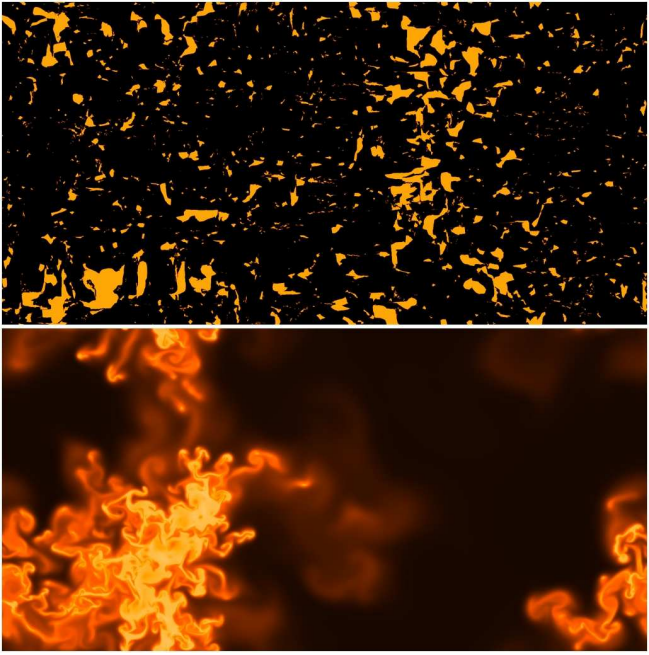,width=\columnwidth}
\caption{Density snapshots at $t \sim 200$ Myr
for the inviscid TI simulations without thermal conduction (top) and with
isotropic uniform conduction of $\mathcal{K}=\mathcal{K}_0$ (bottom).
A logarithmic color scale from $n= 0.28$ $ {\rm cm^{-3}}$ to $n=170$ 
$\rm {cm^{-3}}$ is used.
\label{K_density}}
\end{figure}

Figure \ref{K_density} plots the density at $t\sim200$ Myr in two models:
a calculation without conduction ($\mathcal{K}=0$, top) and one with
$\mathcal{K}=\mathcal{K}_0$ (bottom).  Density structures formed by the
TI are significantly different between the two models.  In the case of
no thermal conduction (top panel), very small dense and cold fragments
begin to form at the beginning of the simulation.  These fragments are
unresolved (1-2 cells in size).  As time proceeds they begin to merge,
and by the end of the calculation at 200 Myr, they dominate the results.
On the other hand, for the simulation with $\mathcal{K}=\mathcal{K}_0$
(bottom panel), resolved density structures begin to emerge at about
20 Myr, and continue to grow into a network of filaments that forms by
about 25 Myr.  By $\sim 33$ Myr, dense cold clouds in thermal equilibrium
form.  The shape and size of the cold clouds evolves due to evaporation
and winds driven from their surfaces by thermal conduction. The smallest 
clouds evident in the density figure have a short lifetime (less then 
$\sim 2$ Myr), as they are quickly dissolved by evaporation, and 
continuously reformed by the evolving density structures.  

\begin{figure}
\epsfig{file=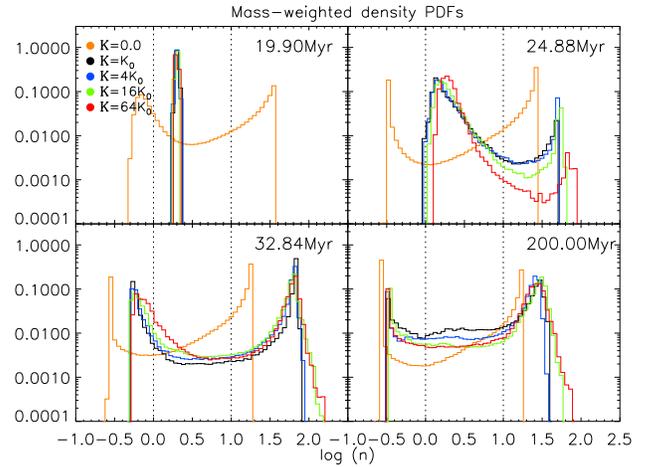,width=\columnwidth}
\caption{Mass weighted density PDFs for inviscid TI
simulations with different values of an isotropic conductivity
at $t \sim 20, 25, 33, 200 \rm~ Myr$.
\label{K_rhoPDF}}
\end{figure}

Figure \ref{K_rhoPDF} shows the mass-weighted density PDFs for the five
models with different level of thermal conductivity. At $t=20 \rm~Myr$,
all the gas in the models with thermal conduction are in the unstable
regime, however in the model without conduction the density is already
segregated into two phases.  Not until $t = 33 \rm~Myr$ do the models
with conduction form density distributions that are clearly separated
into two distinct phases.  After this time, the shape and size of dense 
clouds evolve due to condensation, evaporation, and thermal winds as 
shown in Figure \ref{K_density}, so that there are always some fraction 
of the gas in the interface regions that is in the thermally unstable 
regime.  The mass of gas in the unstable regime {\it increases} as the 
value of the conductivity increases.  Moreover, the highest density also 
{\it increases} with the value of the conductivity.

\begin{figure}
\epsfig{file=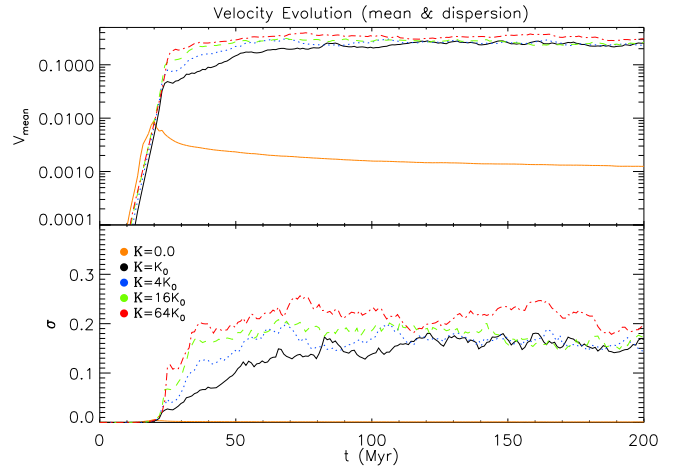,width=\columnwidth}
\caption{Mass weighted mean velocity (top) and velocity dispersion
(bottom) of inviscid TI models with different values of an isotropic
conductivity as a function of time.
\label{K_sigma}}
\end{figure}

In Figure \ref{K_sigma}, we plot the mass-weighted mean velocity, 
defined as $\bar{v} \equiv \sqrt{<\rho v^2>/<\rho^2>}$, and the 
velocity dispersion for each of the five models.  In each case the velocity 
dispersion increases rapidly by TI-induced turbulence. Without thermal 
conduction (orange solid line), the velocity dispersion and average 
velocity attain their maximum values by $t \sim 20 \rm Myr$.  In contrast, 
in the models with thermal conduction, the velocity dispersion and the 
average velocity keep increasing because of the continuous supply of 
unstable gas produced by thermal evaporation from the surfaces of 
dense clouds as discussed in \citet{ino06}. Thus, the model without 
thermal conduction (orange solid line) has the smallest values of both 
the average velocity and velocity dispersion, while the model that 
includes the highest level of conductivity (red dotted line) has the largest.  

\begin{figure}
\epsfig{file=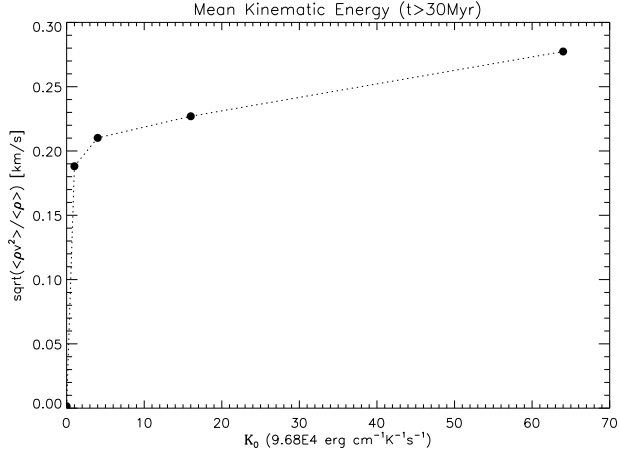,width=\columnwidth}
\caption{The saturated velocity as a function of isotropic thermal conductivity.
\label{K_KEmean}}
\end{figure}

The mass-weighted mean velocity calculated at $t \ge 50$ Myr 
as a function of the thermal conductivity is shown in Figure 
\ref{K_KEmean}.  Clearly the amplitude of the turbulence induced by 
thermal instability increases in proportion to the value of the thermal 
conductivity, at least for $\mathcal{K} > \mathcal{K}_0$.

We draw two conclusions from this study.  First, as has already been 
shown in many recent studies \citep{koy04,pio04}, we conclude it is 
critically important to include explicit thermal conduction in simulations 
of the TI. Figure~\ref{K_density} dramatically demonstrates how 
calculations without conduction lead to unresolved fragments that do 
not evaporate or drive turbulence.  Figure~\ref{K_rhoPDF} shows the 
substantial differences between the PDF of the density in models that 
include conduction, in comparison to a model that does not.  Any study 
of TI-driven turbulence in the ISM that does not include conduction will 
be dominated by errors seeded by the grid.  Second, we conclude it is 
important to adopt the appropriate value for the conductivity.  Adopting 
a larger value in order to resolve the Field length results in 
overestimating the level of turbulence driven by evaporative flows, and 
modifies the PDF of the density.

\subsection{Effect of numerical resolution and viscosity}\label{sec:resolution}

Based on the results from the previous subsection, in the remainder of
this paper we present models in which the conductivity $\mathcal{K}
= \mathcal{K}_0$.  The corresponding Field length is $\lambda_F = 0.12
\rm pc$.  Based on one dimensional simulations, \citet{koy04} have
shown that the Field number $n_{c}=\lambda_{F,min} / \delta x$ should
be greater than 3 to achieve convergence in the number of clouds formed
by the TI, and the maximum Mach number at late time in their simulations.
It is important to confirm that convergence is achieved if the Field length
is resolved in our two dimensional study.  Therefore we have run inviscid
simulations of the TI with isotropic conduction with three different
numerical resolutions: $1024 \times 512$ (low resolution), $2048 \times
1024$ (standard resolution), and $4096 \times 2048$ (high resolution).
All models are computed in a domain of size $2L \times L$, where $L=2$pc.
We have also run models with a constant coefficient of dynamic viscosity
corresponding to a Prandtl number $\rm Pr = 2/3$ at all three resolutions.

\begin{figure}
\plotone{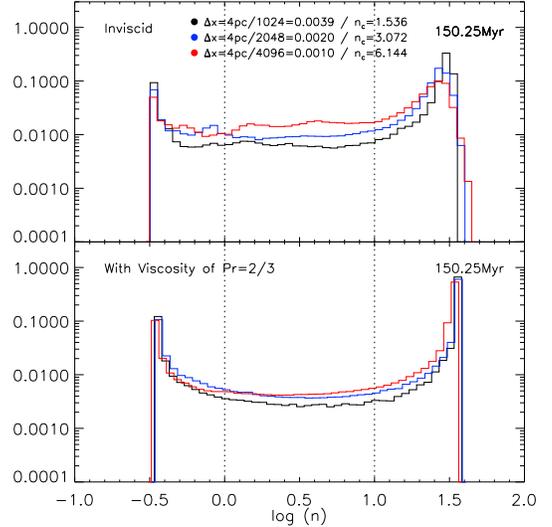}
\caption{Mass weighted density PDFs at $t \sim 150 \rm Myr$ in models
computed with different resolutions.
(Top) Density PDFs of inviscid models.
(Bottom) Density PDFs of models
with explicit viscosity and $\rm Pr=2/3$.
\label{R_study_PDF}}
\end{figure}

Figure~\ref{R_study_PDF} shows the density PDFs at $t \sim 150$ Myr 
both for inviscid models and models with viscosity at different resolutions. 
The PDF of the inviscid calculations {\it (top panel)} are resolution 
dependent, especially for the maximum density and the amount of mass 
in the unstable regime.  The highest resolution inviscid model shows the 
largest value of the maximum density reached in cold gas, and has a 
much larger amount of matter in the unstable regime compared to lower 
resolutions. The latter is a reflection of the fact that matter in the unstable 
regime is contained in dynamical regimes that are driven by thermal 
evaporation, and the properties of these evaporative flows are affected by
viscosity. The bottom panel of Figure~\ref{R_study_PDF} shows that both 
the maximum value of the density and the amount of matter in the 
unstable regime converge with resolution if explicit viscosity is included. 
In this case, the mass-weighted fraction of matter in the unstable regime
is 5.8, 8.2, 9.0 \% for the low, standard, and high resolution models 
respectively.

\begin{figure}
\plotone{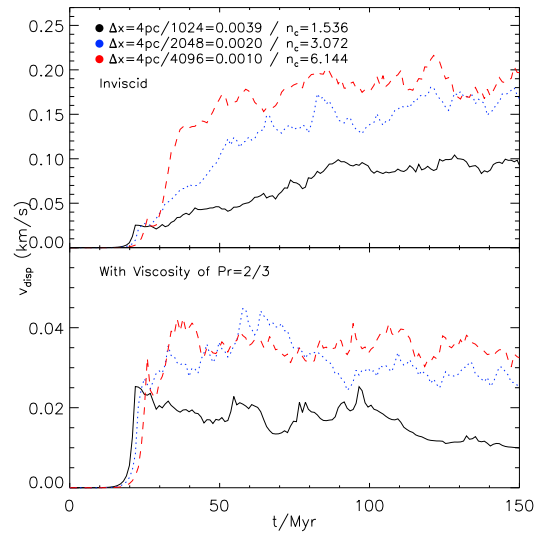}
\caption{Mass weighted velocity dispersion as a function of time at
different numerical resolutions for inviscid calculations (top), and
calculations with explicit
viscosity and $\rm Pr=2/3$.
\label{R_study_sigma}}
\end{figure}

Figure~\ref{R_study_sigma} shows the mass-weighted velocity 
dispersion as a function of time at different numerical resolutions for 
models both with and without explicit viscosity .  In all the models, the
velocity dispersion starts to rapidly increase at around $t\sim20$ Myr,
when dense structures begin to form (see also Figures~\ref{K_density}
and \ref{K_rhoPDF}).  While the velocity dispersion of the inviscid
models continues to increase until $t \sim 50 $ Myr, in the models
with explicit viscosity it remains relatively constant after turbulence
initially develops.  In either case, the velocity dispersion saturates
after $t \sim 50$ Myr, however the mean velocity in the saturated state
of the inviscid models is almost 5 times larger than those that include
explicit viscosity.   Specifically, for $t \ge 50$ Myr the mass-weighted
mean velocity in the inviscid models with low, standard, and high 
resolution is 0.10, 0.19, 0.24 km/s respectively, while models with 
viscosity have mean velocities  that are only 0.017, 0.037, 0.043 km/s. 
This difference is likely due to the fact that the inviscid models have 
$2\--3$ times more matter in the unstable regime, which drives 
correspondingly more powerful evaporative flows. The highest 
resolution model with viscosity has only a 15\% larger mean velocity in 
the saturated state than that of the standard resolution model, which is 
further evidence all quantities are converging.

Based on these convergence tests, we conclude that some properties of 
the nonlinear regime of the TI with isotropic conduction are still resolution
dependent even if the Field length is well resolved if explicit viscosity is 
not included.  For this reason, in studies with anisotropic conduction
presented in the next section, we will present results both with and 
without viscosity.

\section{TI with anisotropic thermal conduction}\label{sec:aniso}

In this section, we investigate the effect of anisotropic conduction on the 
development of the TI by performing magnetohydrodynamics (MHD)
simulations and comparing the results to hydrodynamic simulations of 
the TI with isotropic conduction.  All of the simulations start with our 
fiducial density and pressure, use a conductivity 
$\mathcal{K}=\mathcal{K}_0$, and are computed in a domain of size
$4\times 2 ~\rm pc$ with our standard resolution of $2048 \times 1024$.
This means the Field length is resolved even in the cold phase with at
least 3 cells.

\subsection{Model parameters}
\begin{figure}
\epsscale{1.0}
\epsfig{file=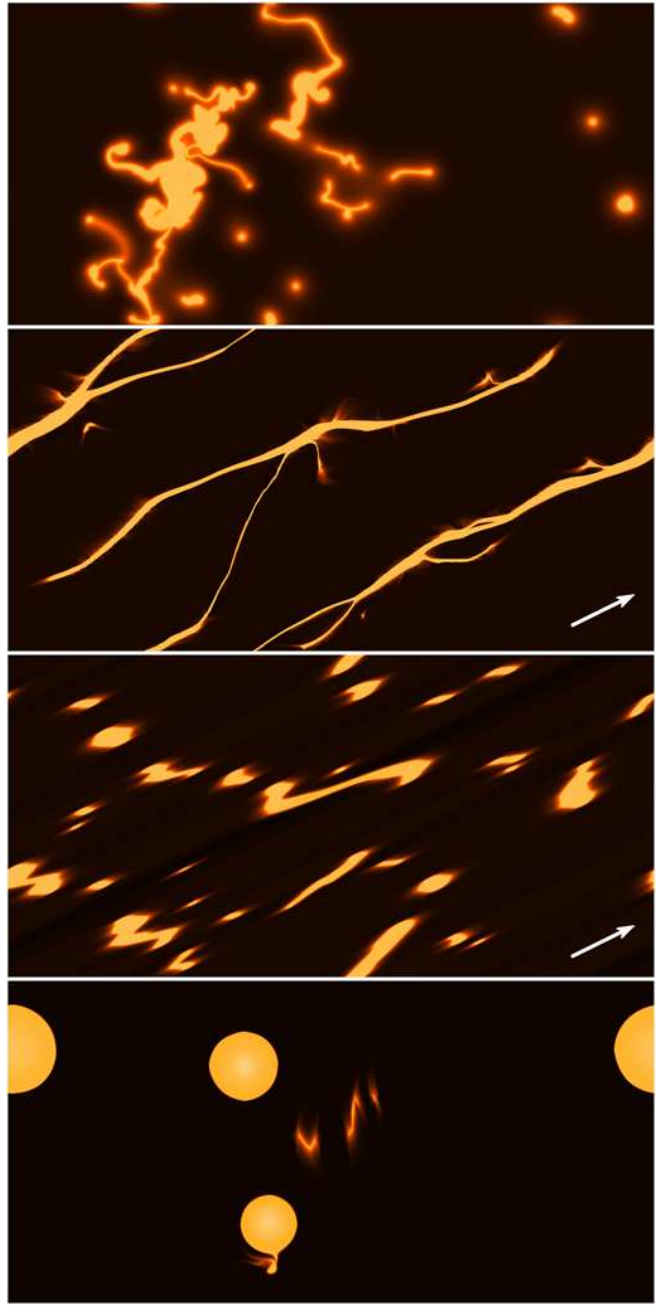,width=\columnwidth}
\caption{Density snapshots at $t \sim 200$ Myr
in TI simulations with 
isotropic conduction (top), anisotropic conduction with weak 
uniform field (second panel), strong uniform field (third panel),
and strong tangled field (bottom) respectively. The arrow shows
the initial direction of the magnetic field.
\label{fig_logd}}
\end{figure}

\begin{table}
   \begin{center}
   \caption{Models with anisotropic conduction}
   \vskip+0.5truecm	
   \begin{tabular}{c|c|c|c|c}\hline\hline
Model & Conduction & Viscosity &Magnetic Field & $ \beta$ \\
  \hline
 1 & hydro/Isotropic &  inviscid & \--&   \-- \cr
2 & mhd/Anisotropic & inviscid & uniform & $10^6$ \cr
3 & mhd/Anisotropic & inviscid & uniform & $1$    \cr
4 & mhd/Anisotropic & inviscid & tangled & $10^6$ \cr
5 & mhd/Anisotropic & inviscid & tangled & $1$    \cr
 {\bf 6} & hydro/Isotropic &   $\rm Pr=2/3$ &  \--&  \--  \cr
 {\bf 7} & mhd/Anisotropic &  $\rm Pr=2/3$ & uniform & $10^6$ \cr
 {\bf 8} & mhd/Anisotropic &  $\rm Pr=2/3$ & uniform & $1$    \cr
 9       & mhd/Anisotropic &  $\rm Pr=2/3$ & tangled & $10^6$ \cr
 {\bf 10}& mhd/Anisotropic &  $\rm Pr=2/3$ & tangled & $1$    \cr
  \hline\hline
   \end{tabular}
   \end{center}
   \label{tab2}
 \end{table}

\begin{figure}
\epsfig{file=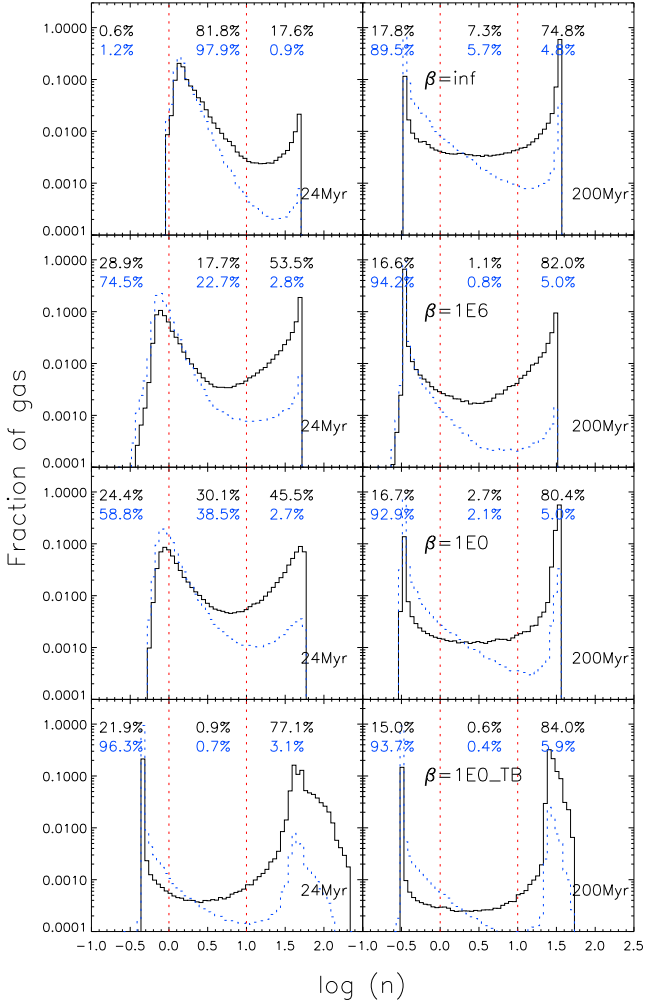,width=\columnwidth}
\caption{Mass (black solid line) and volume (blue dotted line) weighted 
density PDFs for models 6, 7, 8 and 10 at $t \sim 24$ 
and 200 Myr. The percentages of gas in each phase by mass are noted
in black, and the proportions by volume are shown in blue.
\label{fig7}}
\end{figure}

\begin{figure}
\epsfig{file=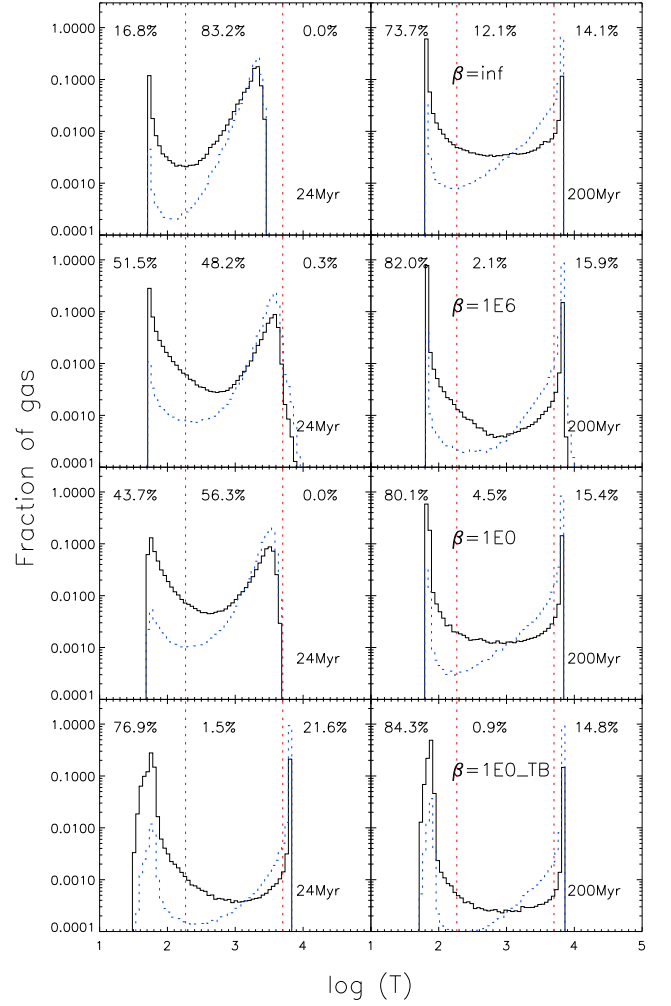,width=\columnwidth}
\caption{Mass (black solid line) and volume (blue dotted line) weighted 
temperature PDFs for models 6, 7, 8 and 10 at $t \sim 24$ and 200 Myr. 
The percentages of gas in each phase by mass are shown.
\label{fig8}}
\end{figure}

We have performed a series of simulations in order to explore both the 
effect of varying the magnetic field geometry and strength on the nonlinear 
outcome of the TI.  To explore the effect of geometry, we have computed 
models with an initially uniform magnetic field inclined along the diagonal 
of the domain (to eliminate potential artifacts that might be produced by 
aligning the field with the grid), as well as a random tangled field 
generated using the technique described below.  In both cases, we study 
two different initial magnetic field strengths corresponding to 
$\beta = P_{gas}/P_{mag} = [10^6,1]$. For our fiducial initial conditions, 
these field strengths are $B = [0.00323,3.23]$ $\mu G$.

To generate the tangled magnetic field, we initialize a vector potential 
${\bf A} = (A_x, A_y, A_z)$ which has only one non-zero component that 
is given by a Fourier power spectrum of the form
\begin{equation}
A_z = |\delta {\rm A}_{k} | \propto k^{-11/6}
\end{equation}
with amplitudes that follow a Gaussian random distribution for all 
wavenumbers $k$ in the range $2\pi/(L/2) \le k \le 2\pi/\delta x$. The 
magnetic field is then ${\bf B} = \nabla \times {\bf A}$, and is
normalized so that the volume averaged magnetic energy gives the
appropriate value of $\beta$.

Table 2 summarizes the properties of our runs. They include two models 
with isotropic conduction (one with and one without viscosity), as well as 
8 runs with anisotropic conduction with different magnetic field strengths 
and geometry.   We study the TI with anisotropic conduction using both 
models that are inviscid, and models that include explicit viscosity with 
${\rm Pr} = 2/3$.

In this paper, we present four of our simulations:  a hydrodynamic model
with the isotropic conduction in an unmagnetized medium (Model 6) 
and MHD models with the anisotropic conduction in magnetized medium 
(Model 7, 8, and 10). For detailed study on the formation of clouds by
the TI with isotropic conduction or without thermal conduction
in a magnetized medium, we refer to recent papers: 
\citet{hen08,hei08,ino09,ban09}.

\subsection{Evolution of the density}

Figure \ref{fig_logd} shows snapshots of the density at $t \sim 200$ Myr 
from four of our simulations: a hydrodynamic model with isotropic 
conduction, MHD models with anisotropic conduction and both a weak 
and strong uniform field, and an MHD model with anisotropic 
conduction and a strong tangled field.  In each case, the TI results in 
cold regions that are more than an order of magnitude denser than their
surroundings, however, the distribution and evolution of the cloud blobs
are significantly different in each case.

For the hydrodynamic model with isotropic conduction (top), a network
of filaments forms at about 25 Myr connecting the regions of highest
density. By $\sim 33$ Myr, dense cold clouds begin to form.  Clouds with 
a radius smaller than the Field length $\lambda_F$ are destroyed by 
thermal conduction.  Evaporative flows form around larger cold clouds, 
they are evident as the ``fuzzy" edges of dense regions in 
Figure~\ref{fig_logd}. The evaporative flows cause the dense clouds and 
filaments both to merge and fragment in random motions.

For the MHD model with anisotropic conduction and either a weak or
strong uniform magnetic field, the regions of cooled gas tend to form
thin filaments parallel to the direction of the field lines.  In the linear 
regime, the preference for filamentation along the fields can be
explained by the variation of the Field length with angle with respect
to the direction of field.  Perpendicular to the field, the Field length
is very small, so very short wavelength perturbations are unstable.
Parallel to the field, the Field length is much longer, so only longer
wavelength modes grow.  As the perturbations grow nonlinear, 
conduction tends to enforce isothermality along magnetic field lines, 
leading to long filaments.  In the case of strong fields, only motions along 
field lines are allowed, and moreover magnetic pressure provides some 
support in dense regions, resulting in fragments aligned with the field 
lines as evident in the third panel of Figure \ref{fig_logd}.

For the model with a strong tangled magnetic field (bottom panel),
spherical dense clouds quickly develop at the centers of regions of 
closed field lines (the local minimum and maximum of the vector 
potential). The topology of the closed field lines in these regions means 
they become thermally isolated from the rest of the domain, and therefore
neither evaporate nor evolve further.  Small clouds originally formed in 
regions of open field lines eventually evaporate along the magnetic field 
lines and disappear.

\subsection{Density and temperature statistics}

Figure \ref{fig7} shows the mass and volume weighted density PDFs at
both 24 (left column) and 200 (right column) Myr for the four different 
models shown in Figure~\ref{fig_logd}; from top to bottom the panels are 
for isotropic conduction, anisotropic conduction with weak and strong 
uniform field, and anisotropic conduction with a strong tangled field 
respectively.  As was shown in section \ref{sec:coneffect}, thermal 
conduction can reduce the growth rate of the TI.  However, models with 
anisotropic conduction develop density distributions with distinct peaks 
in the dense (cold) and diffuse (hot) phases earlier than the model with 
isotropic conduction.  Similarly, by the end of the simulation, the 
hydrodynamic model with isotropic conduction has a larger amounts of 
mass in the unstable regime  compared to the models with anisotropic 
conduction.  This interpret this result as due to the suppression of 
evaporative flows from the surfaces of dense clouds by the magnetic field 
and anisotropic conduction.  As shown earlier, material in the unstable 
regime is contained in evaporative flows (the fuzzy edges of the clouds 
seen in the top panel of Figure~\ref{fig_logd}). With anisotropic 
conduction, thermal conduction can only occur from the surfaces of the 
clouds where the direction of the magnetic field is normal.  For the 
elongated and filamentary clouds formed with anisotropic conduction, 
only a very small fraction of its surface area is subject to evaporation, 
resulting in very little unstable gas.  For the same reason, the fraction of 
high density gas (cold phase) is slightly larger in the anisotropic models 
compared to that of the isotropic case.

Probably the most striking feature of the density PDFs are the distinct
peaks representing the two phase medium at 200 Myr.  The mass and 
volume weighted temperature PDFs of these four models shown in 
Figure~\ref{fig8}, reflect this structure: most of the matter is either in the 
cold or hot phases.  Very little is contained in the unstable regime at 
intermediate temperatures, and the fraction of gas in the unstable regime 
decreases in the anisotropic conduction models.

\subsection{Evolution of the kinetic energy}

\begin{figure}
\epsfig{file=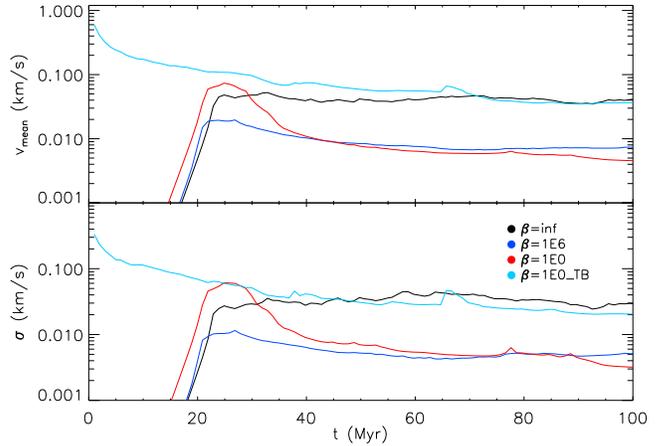,width=1.0\columnwidth}
\caption{Mass weighted mean velocity ({\it top}) and velocity dispersion 
({\it bottom}) as a function of time from models with hydrodynamic
conduction ($\beta = \inf$),  1, 6, 
7, 8.
\label{mhd_velo}}
\end{figure}

As shown in section \ref{sec:coneffect}, the TI drives evaporative flows 
that produce turbulence.  Figure~\ref{mhd_velo} compares the mass
weighted mean velocity and velocity dispersion in models that include
isotropic conduction, anisotropic conduction with a weak uniform field
or strong uniform field, and a strong tangled field.  The evolution of
these quantities is a direct measure of the strength of evaporative flows.

Except in the case of the tangled field, the mean velocity and dispersion
grow initially, until dense clouds form and the TI saturates around 25 Myr. 
In the case of the strong tangled field, there are unbalanced Lorentz 
forces in the initial conditions that cause motions immediately. These 
decay away as the TI saturates.  Thus, the time evolution of the velocity 
and its dispersion is not a good measure of TI driven turbulence in this 
case.  For the other two models with anisotropic conduction (the strong 
and weak uniform field cases), the saturation amplitude of the mean 
velocity and dispersion at late times are about $5\--10$ times lower in the 
models with anisotropic conduction.  This is clear evidence that 
evaporative flows from the surfaces of the dense clouds is suppressed by 
the magnetic field and anisotropic conduction. This is because 
evaporation can only occur when the field line is normal to the interface.
For most of the surface of long filaments, the field is parallel to the 
interface.  This, only a very small region at the end of the filament can 
produce evaporative flows.

\subsection{Amplification of the magnetic field}\label{sec:bfield}

\begin{figure}
\epsfig{file=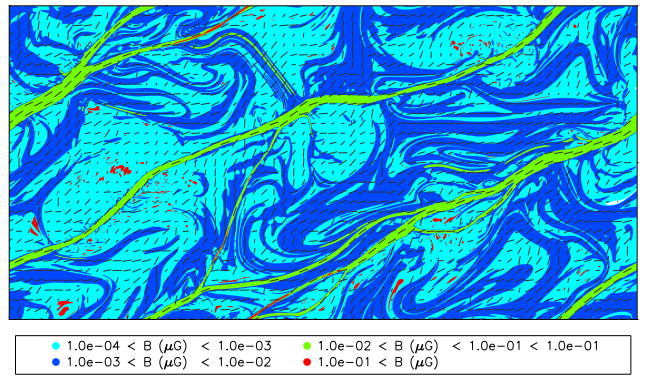,width=1.0\columnwidth}
\caption{Contour plot of the magnetic field strength (color) and magnetic 
field lines at $t \sim 200$ Myr in TI simulation with anisotropic conduction 
and with weak uniform field (Model 7).
\label{mhd_bfield}}
\end{figure}

In Figure~\ref{mhd_bfield}, we show an image of the magnetic field
strength overlaid with line segments to show the direction of the
magnetic field lines at $t \sim 200$ Myr in a TI simulation with
anisotropic conduction and with weak uniform field (Model 7). The
magnetic field is initially uniform with a strength of 0.00323 $\mu$G. As 
material accumulates into the filaments by TI, the magnetic fields are 
compressed, resulting in strong fields that are parallel to the direction of 
the filament.   Outside of the filaments, the magnetic field is folded and 
twisted by vortical fluid motions.  We find the compression and twisting 
can amplify the field up to 0.3 $\mu$G, i.e., by a factor of 100, as shown
in \citet{ino07}.  In full three-dimensions, the geometrical compression 
during the formation of filaments is likely to be larger than the 
two-dimensional case studied here, thus the factor of 100 amplification
may be a lower bound.  Note the regions where the magnetic field is 
strongest trace the cold hydrogen filaments.

\section{Summary}\label{sec:Summary}

In this work, we have studied the nonlinear regime of the thermal
instability (TI) and the effect of anisotropic thermal conduction by
performing two-dimensional hydrodynamical and magnetohydrodynamic
simulations incorporating radiative cooling and heating.  Our main
results can be summarized as the following:

1.  As found in previous studies, it is crucial to include explicit thermal
conduction in numerical studies of the TI so that the Field length is
resolved, in order to prevent artificial fragmentation
driven by numerical noise at the grid scale (e.g. Figure~\ref{K_density}).

2. The amplitude of the thermal conductivity controls the rate of
evaporation from the surfaces of dense clouds formed by the TI, and
therefore strongly affects the amplitude of turbulent motions induced
by the TI (e.g. Figure~\ref{K_KEmean}).

3. Even when the Field length is resolved, explicit viscosity must
be included to obtain numerical convergence of some quantities, for
example the amplitude of turbulent motions driven by evaporative
flows (e.g. Figure~\ref{R_study_sigma}).

4.  Although the statistics of the density and temperature are not
strongly affected by anisotropic conduction, the geometry of structures
formed by the TI are quite different.   With anisotropic conduction and
a uniform magnetic field, the TI saturates as long thin filaments of
dense gas aligned with the field.  In a tangled field, spherical clouds
are formed in regions of closed field lines (e.g. Figure~\ref{fig_logd}).

5.  The combination of anisotropic conduction and MHD strongly 
suppresses the rate of evaporation of cold gas from the surfaces of 
dense structure in regions where the field is parallel to the interface.  
This reduces the amplitude of turbulence driven by the TI 
(e.g. Figure~\ref{mhd_velo}).

These results have a number of implications for observations of cold
neutral gas in the ISM.  In particular,  the thin filaments along the
magnetic field in the weak uniform magnetic field case agree well with
recent observations of the Riegel-Crutcher cloud conducted by 
\citet{mcc06}. This neutral hydrogen (HI) cloud lies on the edge of the Local
Bubble \citep{cru84} filled with a hot and diffuse gas where the anisotropic
conduction can be expected. \citet{mcc06} found a network of dozens of 
hairlike filaments of cold hydrogen with widths of less than $\sim 0.1$ pc 
and up to 17 pc long. They also have found that the filaments are aligned 
with the magnetic field of the cloud which agrees well with our results. 
They also calculated the magnetic field strength by using the 
Chandrasekhar-Fermi method, finding $\sim 60 \mu$G.  We find that 
compression and twisting of field during the formation of filaments by the 
TI with anisotropic conduction can amplify it by up to a factor of 100 in 
two-dimensions.  In three-dimensions, the amplification is likely to be larger. 
This may be enough to explain the observed field.

Recently, \citet{sha10} have reported a study of the TI with anisotropic 
thermal conduction in the hot X-ray emitting plasma in clusters of 
galaxies. The heating and cooling processes in this regime are very 
different than those in the ISM studied here (equations \ref{eq:E_heat} 
and \ref{eq:E_cool} respectively), in particular there are no stable 
phases in the cooling curve they adopt, so that magnetic pressure sets 
the only limit on how cold and dense the gas becomes.  Nonetheless, 
the structure of density condensations in this case are very similar to 
our results: filaments of cold gas along magnetic field lines.

Our results show that in the case of anisotropic conduction, the
geometry of the magnetic field with respect to the interface between
cold and hot phases is very important.  Only if the field is normal to the 
interface can thermal conduction drive evaporation and outflows. The 
inclusion of anisotropic conduction could have important implications
for the structure of interfaces, and the interpretation of observations of 
these regions \citep{ino06,sto09,sto10}.

There are a number of limitations to our work that should be addressed
in future investigations.  Firstly, the ISM is highly turbulent \citep{hei03}, 
with typical turbulent velocities approximately 7 $\rm km s^{-1}$ 
\citep{hei03,moh04}, i.e. more than 100 times larger than those produced 
by the TI with isotropic conduction and viscosity. In the traditional picture 
of the ISM, supernova driven turbulence leads to a hot, diffuse third phase 
\citep{cox74,mck77}.  More recently, \citep{pio04} have considered the 
interaction of turbulence driven by the MRI and TI. We have not 
considered the effect of externally forced turbulence on the TI in this work, 
however this would be a productive direction for study in the future.

Secondly, in this work we have adopted a constant conductivity
$\mathcal{K}$.  However, in reality the conductivity is a function of the 
temperature \citet{par53}, so that the rate of thermal conduction
decreases as the gas cools. The amplitude of the conductivity can vary
by 2 orders of magnitude between warm and cold phases at the end of
our typical simulation at $t\sim 100$~$\rm Myr$.  As discussed in Section
\ref{sec:coneffect}, the value of the conductivity can affect the rate of 
evaporation from dense clouds, and therefore the kinetic energy of 
turbulence driven by the TI, thus assuming a constant value may alter
the result.  Using  a realistic temperature dependent conductivity is
challenging, since it would decrease the minimum value of Field length
and require much higher resolution.  Since the geometry of the magnetic
field limits the amount of evaporation to a very small surface area at the 
ends of the filaments in the case of anisotropic conduction, we do not 
expect the use of a more realistic conductivity to produce qualitative 
changes in our result.  Nonetheless, more realistic studies which use 
temperature dependent conductivities would be fruitful.

Finally, this study has considered flows in only two dimensions. In full 
three dimensions, the amplification of the magnetic field due to 
geometrical compression into filaments may be larger, and the 
turbulence driven by evaporative flows may be of a different character.
Fully three-dimensional simulations of the TI with anisotropic conduction
would also be interesting for future studies.

\acknowledgments
We thank M. Ryan Joung, Eve Ostriker, Bruce T. Draine and Prateek 
Sharma for useful discussions. We thank the anonymous referee for the 
constructive comments. This work used computational facilities provided by
the Princeton Institute for Computational Science and Engineering. We
acknowledge support from a grant funded through the NASA ATP.

\end{document}